# Is the surface fluid during surface melting – a study of the surface shear modulus of solid Gallium

Almog Danzig[1], Ori Scaly and Emil Polturak[2]

Physics Faculty, Technion Institute of Technology, 32000 Haifa, Israel

Melting of 3D solids is often preceded by a melting of their surface, a distinct process which begins at a temperature lower than $T_M$. Until now, surface melting was investigated mostly by diffraction or other non-contact techniques, revealing how the surface becomes progressively disordered with temperature. We designed a method to measure an effective shear modulus of the surface, another property which should change at melting. This property was not measured before. We applied this method to surface melting of Gallium, and found that this surface shear modulus vanishes abruptly near the onset of premelting, about 9K below $T_M$.

Historically, melting is one of the most extensively studied physical phenomena[1–4]. Potential mechanisms of melting include enhancement of atomic vibrations[4] and surface melting [5–15]. Intuitively, melting should begin in a weak region of the solid, such as the surface or a grain boundary, from where the liquid phase will propagate into the bulk. For that reason, surface melting is the most likely precursor of melting. It is often called premelting, as it takes place at a lower temperature than melting of the bulk.

The attributes of bulk melting include a loss of the crystallographic order, a latent heat, vanishing of the shear modulus and a discontinuous change of the density. All those changes occur simultaneously at the melting point. Surface melting on the other hand is a gradual process of disordering of the surface layer with temperature. It was first detected at the surface of lead[5–7]. Other solids showing surface melting include Ice [9,16,17], Gallium [15,18], Methane [19,20], Ag[21,22], Al [23–27], Cu[28], Si(111)[12,29] and Au (111) [30,31]. Simulations showed that surface melting starts at the least dense surface [32,33]. Until now, surface melting was detected mostly by monitoring the structural properties of the surface as a function of temperature. A gradual loss of the crystalline order in the topmost atomic layers was interpreted as surface melting. The reason why so far only structural probes were used is twofold. First, scattering from the surface is a non-contact method which can be applied to a variety of materials and over a wide range of temperatures. Second, scattering techniques can probe only a few atomic layers nearest to the surface. With other methods, for example specific heat, it is difficult to separate the contribution of the surface from that of the bulk.

Regarding the shear modulus, in the bulk it decreases discontinuously to zero at the melting point[34,35]. When applying shear stress at the surface of a macroscopic solid, the elastic response is influenced by the bulk material below the surface. Under normal conditions, it is not possible to separate the response of the surface from that of the bulk. Consequently, until now there are no experimental measurements of the response of the surface to shear. Vanishing of the shear modulus is the only mechanical indication of melting. We therefore decided to try to measure the effective response of the surface layer to shear. We focused on the temperature range where premelting was observed. We chose to work with Gallium, because it undergoes a surface melting transition which is well documented[15,18].

[1] Almog.Danzig@gmail.com
[2] Emilp@physics.technion.ac.il

We first set up a qualitative experiment that allows us to apply shear stress to the surface and detect a surface strain. To this end, we dispersed small Iron (Fe) particles, with a typical size of 1-5 $\mu m$, on the surface of a polycrystalline Gallium. A typical particle on the surface can be seen in a SEM image in the inset of Figure 1. Inspection of many such images shows that the Fe particles do not penetrate into the bulk and are attached to the surface only by static friction. We apply a weak alternating force on these particles by moving a small permanent magnet back and forth beside the Ga surface (see inset of Figure 1). The tangential component of this force acts on the Fe particles in parallel to the surface. The lateral displacement of the Fe particles is viewed under a microscope and recorded by a camera. As long as the magnetic force is smaller than the typical static friction force, the Fe particles do not move <u>relative</u> to the Ga surface, but they do exert a force having a tangential component on the surface. Under these conditions, any displacement of these particles represents the local strain of the Ga surface. After image processing, we can determine an average displacement of the surface with a resolution of a few nanometers. In the experiment, we measured this displacement as a function of temperature of the Ga. The results are shown in Figure 1. Below 294.2K, the displacement is close to zero. Above 294.2K, we observed a dramatic increase of the displacement, by more than an order of magnitude. The displacement becomes so large that it can actually be observed visually under the microscope without any image processing. The onset temperature of ~294.2K is slightly higher than 293.15K, the temperature where Rühm[15] observed premelting. Two transitions can be seen in our measurement: enhancement of the displacement by more than 2 orders of magnitude at 294.2K, near the premelting transition, and another enhancement by factor of 4 at the bulk melting temperature. At 294.2K the solid appears to undergo a kind of relaxation process involving reorientation of solid grains on the surface. This change could result from stress release on a local scale. The surface shows no additional changes up to the bulk melting temperature, where it becomes smooth with a mirror-like appearance. These observations gave us confidence that changes of the shear properties of the surface can be detected using our technique and connected directly with the surface melting of Gallium.

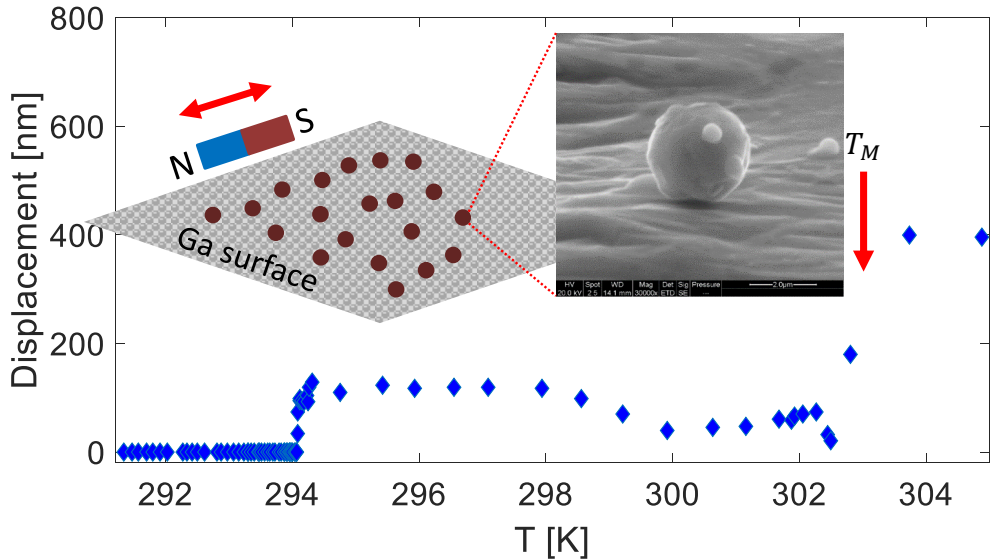

*Figure 1: Temperature dependence of the displacement of Iron particles dispersed on the surface of Ga under a magnetic force. The amplitude of the force is independent of temperature. The inset shows the experimental layout as well as an SEM image of a typical particle. The displacement of the particles is effectively zero below 294.2K and increases sharply at this temperature. A second enhancement of the displacement occurs at the bulk melting temperature.*

In the next step, we developed a technique which gave us better control over the force we apply on the Ga surface. This technique enabled us to do more quantitative measurements. The experimental arrangement is shown in the inset of Figure 2. A small metal disk (2mm diameter) is soldered to the surface of the Ga slab using the Ga itself as the solder. The disk is connected to a piezoelectric actuator by a thin metal rod. The piezo actuator applies an alternating force on the disk, in a direction parallel to the surface. The disk in turn exerts shear stress on the surface. The absolute value of the applied stress is determined in a separate calibration, using a strain gauge installed inside the actuator. In this setup we have a continuous measurement of the applied force. The metal disk and the Gallium surface around it are observed under a microscope and photographed with and without stress. Differential image analysis of these photographs yields a map of the local displacement at each point of the surface with a few nm resolution. This experiment is repeated as a function of temperature.

Qualitatively, we see that below a temperature of 294.2K the entire surface of the Gallium responds to the motion of the metal disk. The response gets weaker as one moves away from the disk. Above 294.2K the Gallium surface becomes decoupled from the motion of the disk. Only the boundary layer very near the disk continues to move with the disk. The transition temperature does not depend on the motion amplitude of the metal disk. This confirms that the premelting is not a result of friction between the Ga surface and the disk. The transition temperature is the same as in our first experiment.

In Figure 2 we present the motion amplitude of the metal disk measured by a strain gauge, and the magnitude of the force applied on the metal disk as a function of temperature. One can see that at the transition temperature, the motion amplitude increases sharply while the magnitude of the applied force decreases. The amplitude increases to the same value which we measured with a free metal disk. This value, constrained by the piezo system, represents the case where the surface has zero shear resistance.

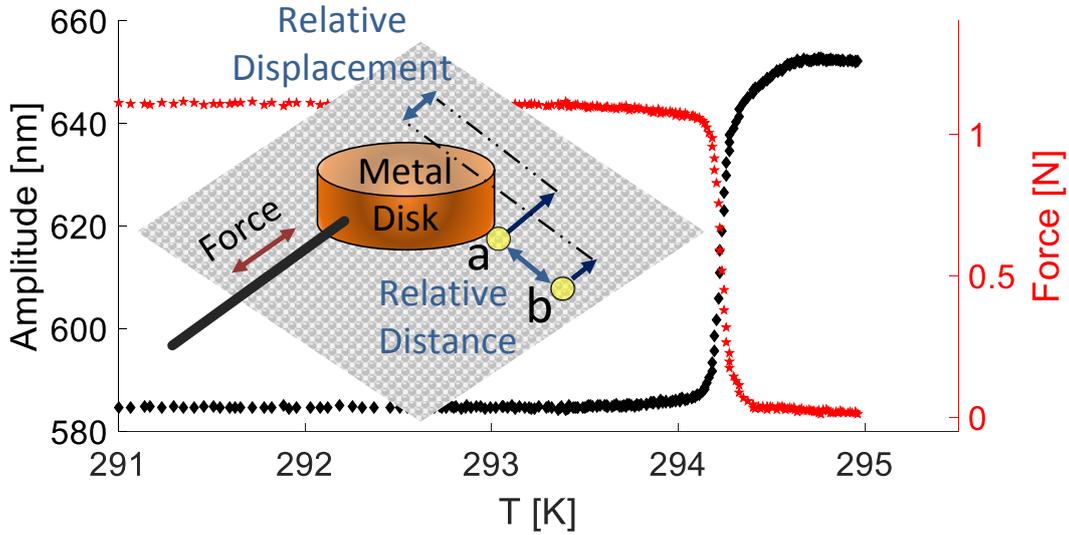

*Figure 2: Motion amplitude of the disk (black symbols) vs. temperature under an alternating driving force. The red symbols show the magnitude of the force required to sustain the motion. The inset shows the layout of the experiment. The actuator applies force on the disk attached to the Ga surface. We measure optically the displacement of two small regions on the surface: Region **a** which is close to the disk and region **b** which is far from the disk.*

Since we are able to do quantitative measurements, it is interesting to determine the magnitude of the effective surface shear modulus and compare it with that of the bulk. We define the ratio between stress and strain as an effective surface shear modulus ($G_s$) as follows:

*1 Equation*

$$G_s = \frac{Force/Metal\ disk\ area}{Relative\ displacement/Relative\ Distance}$$

The various quantities used in Equation 1 are illustrated in the inset of Figure 2. With $G_s$, the relative distance and displacement are defined on the surface. In contrast, with the bulk shear modulus, the relative distance and displacement are defined between the top and bottom of the sample. Referring to the layout shown in Figure 2, we measure the relative displacement by comparing images of two adjacent regions on the surface. The displacement of region **b** is smaller than that of **a**. While the surface is solid, the displacement of region **a** is larger than that of region **b** in proportion to $G_s$ and the relative distance between **a** and **b**. The strain is independent of the choice of **a** and **b**. Once premelting took place, the displacement of region **a** increases and the displacement of region **b** decreases to zero.

We measured the displacement using image processing based on a cross correlation function. The relative displacement is the difference between the displacements of the two regions. After averaging, this method yields the relative displacement with a few nanometer resolution. Measurements of the shear modulus of bulk Ga were done using a standard method for bulk solids.

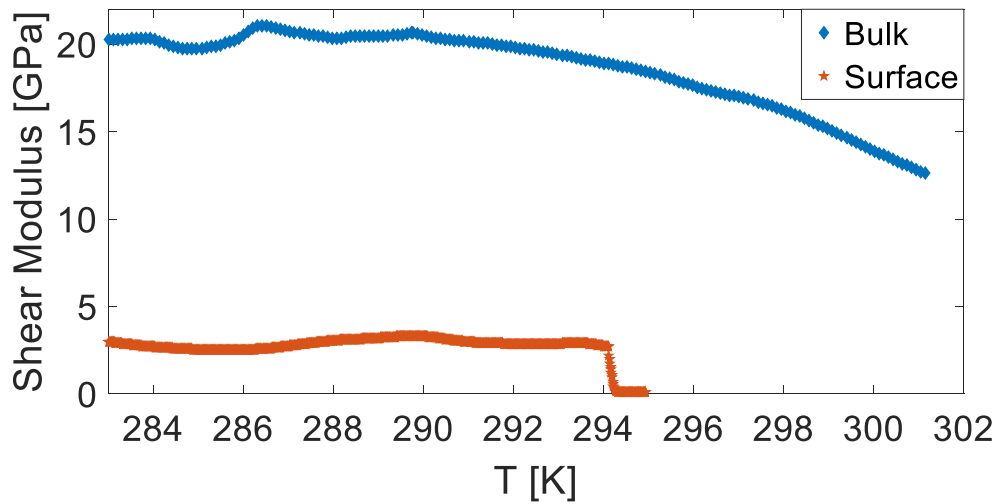

*Figure 3: The shear modulus of bulk Ga and the effective shear modulus of the surface of Ga vs. temperature. The effective surface shear modulus vanishes abruptly at 294.2K.*

In Figure 3 we compare the shear moduli of the surface and of the bulk. Away from melting, $G_s$ is about 5 times smaller than that of the bulk. $G_s$ vanishes abruptly at 294.2K. There is a large temperature difference between the surface and bulk transitions. Both our results and those of Rühm[15] show that premelting of Gallium is distinct from bulk melting.

The temperature at which $G_s$ vanishes is about 1K higher than the onset temperature given by Rühm, et al.[15]. In their experiment, the thickness of the disordered layer increased smoothly starting from zero at 293.15K up to ~7 atomic layers at the bulk melting temperature (~303K). Rühm[15] determined the thickness of the liquid layer by analyzing his data using the theory of Lipowsky[36]. According to this estimate, at the temperature where $G_s$ vanishes (294.2K), the thickness of the liquid is ~0.2 of an atomic layer. One possible explanation as to how our experiment fits with Rühm's is that the disorder reaches some percolation threshold. Beyond this threshold, the layer may lose its resistance to shear and becomes fluid-like. In the presence of fluid, solid grains can slide past each other and the shear modulus would vanish. On a macroscopic scale we will see fluid-like behavior although each grain is still a solid.

To summarize, we measured the effective shear modulus of the surface of Ga near melting. The modulus vanishes at T~294.2K, about 1K above the onset of premelting[15]. In addition, the solid appears to undergo a morphological change which looks like a reorientation of solid grains on the surface. This change could result from stress release on a local scale. Our measurement provide the first observation of surface melting from the mechanical perspective, i.e. onset of fluidity.

## Acknowledgment

We thank S. Lipson and G. Koren for helpful advice. We are grateful to S. Hoyda, L. Yumin, and the Physics Faculty Workshop for technical help. This work was supported in part by the Israel Science Foundation.